# Analysis of the Transport Process Providing Spin Injection through an Fe/AlGaAs Schottky Barrier

A.T. Hanbicki,[a)] O.M.J. van 't Erve, R. Magno, G. Kioseoglou, C.H. Li,[b)] and B.T. Jonker
*Naval Research Laboratory, Washington, DC 20375*

G. Itskos, R. Mallory, M. Yasar and A. Petrou
*State University of New York at Buffalo, Buffalo, NY 14260*

Electron spin polarizations of 32% are obtained in a GaAs quantum well via electrical injection through a reverse-biased Fe/AlGaAs Schottky contact. An analysis of the transport data using the Rowell criteria demonstrates that single step tunneling is the dominant transport mechanism. The current-voltage data show a clear zero-bias anomaly and phonon signatures corresponding to the GaAs-like and AlAs-like longitudinal-optical phonon modes of the AlGaAs barrier, providing further evidence for tunneling. These results provide experimental confirmation of several theoretical analyses indicating that tunneling enables significant spin injection from a metal into a semiconductor.

Significant effort has been made to incorporate ferromagnetic metals into semiconductor spintronic devices because they offer high Curie temperatures and a ready source of spin polarized electrons. However, theory indicates that only very small spin injection effects (< 0.1%) can be expected for intimate metal/semiconductor contacts if transport across the interface occurs via a process that can be accurately described using a classical diffusion equation model.[1,2] It has been shown that this obstacle can be overcome if the interface resistance dominates, such as when the carriers are injected from the metal into the semiconductor by tunneling through a barrier.[3,4,5,6] Indeed, several recent experimental efforts have reported successful spin injection into semiconductor heterostructures from ferromagnetic metals using a variety of tunnel barriers, including Schottky contacts,[7,8] thin metal oxides,[9,10] and AlAs.[11]

Although injection of spin polarized carriers was attributed to tunneling in each of these studies, no conclusive evidence was presented to indicate that tunneling was the dominant transport mechanism. In contrast, abundant evidence has been provided in the metal / insulator / metal (MIM) tunnel junction community. For instance, it has recently been shown that issues such as pinholes in a metal oxide, a chronic problem with these types of heterostructures, can not easily be ruled out.[12,13,14] Therefore, we present here a detailed study of tailored Fe/AlGaAs Schottky barriers where significant spin injection is demonstrated to occur and analyze the transport process. Application of the "Rowell criteria" for tunneling[12,15] and observation of phonon signatures in the low temperature conductance spectra conclusively demonstrate that tunneling is the dominant transport mechanism.

A Schottky barrier provides a natural tunnel barrier between a metal contact and a semiconductor, obviates the need for a discrete layer, and is already a routine ingredient in semiconductor device technology. Injection of electrons from metal to semiconductor occurs under reverse bias, but is usually minimal due to a wide (~ 1000 Å) depletion region in the semiconductor. The use of a thin, heavily doped surface region in the Fe/AlGaAs/GaAs heterostructures reported here reduces the depletion width as well as the effective barrier height, significantly enhancing the probability for tunneling.[16,17] Samples were grown by molecular beam epitaxy (MBE) using interconnected growth chambers. The semiconductor heterostructure consisted of either an $n$-AlGaAs epilayer on an $n$-type GaAs(001) substrate, or a quantum well (QW) structure of 850 Å $n$-$Al_{0.1}Ga_{0.9}As$ / 100 Å undoped GaAs / 500 Å $p$-$Al_{0.3}Ga_{0.7}As$ / $p$-GaAs buffer layer on a $p$-GaAs(001) substrate. The latter structure forms a spin-polarized light emitting diode (spin-LED) which was used to determine the spin polarization of the electrons in the GaAs QW. The top

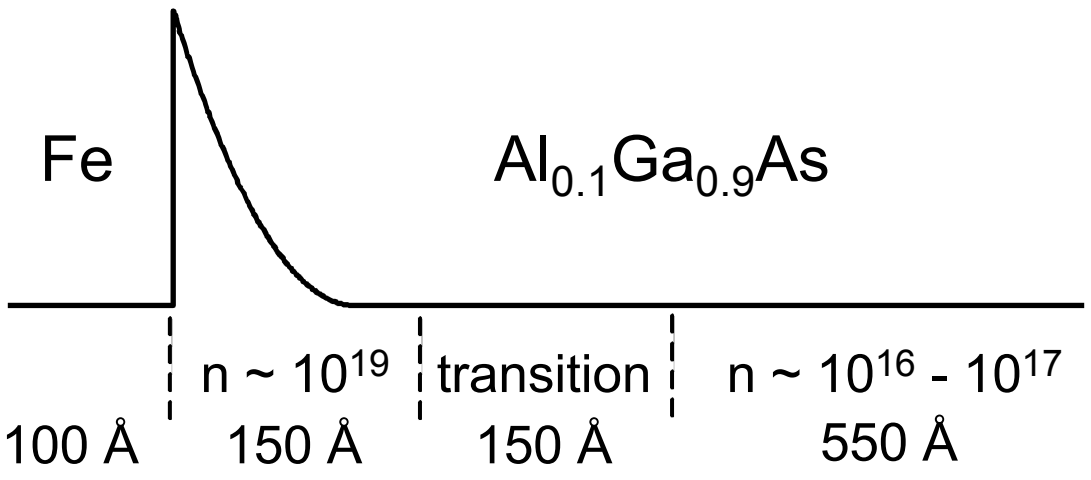

FIG. 1. Schematic flat band diagram of the interface used here.

150 Å of $n$-type $Al_{0.1}Ga_{0.9}As$ was doped $n$=1x10$^{19}$ cm$^{-3}$ to minimize the depletion width. This was followed by a 150 Å transition region, while the rest was $n$=1x10$^{16}$ cm$^{-3}$ with a 100 Å dopant setback at the QW. A 100 Å thick Fe(001) film was grown in the second MBE chamber with the substrate at 10-15˚C to minimize potential intermixing. Additional details of the growth are described elsewhere.[8] Figure 1 shows the doping and a schematic flat band diagram at the Fe/AlGaAs interface. The resulting Schottky contact has a narrow depletion width and forms a triangular shaped tunnel barrier, enabling spin polarized electrons to tunnel from the Fe into the semiconductor under reverse bias.

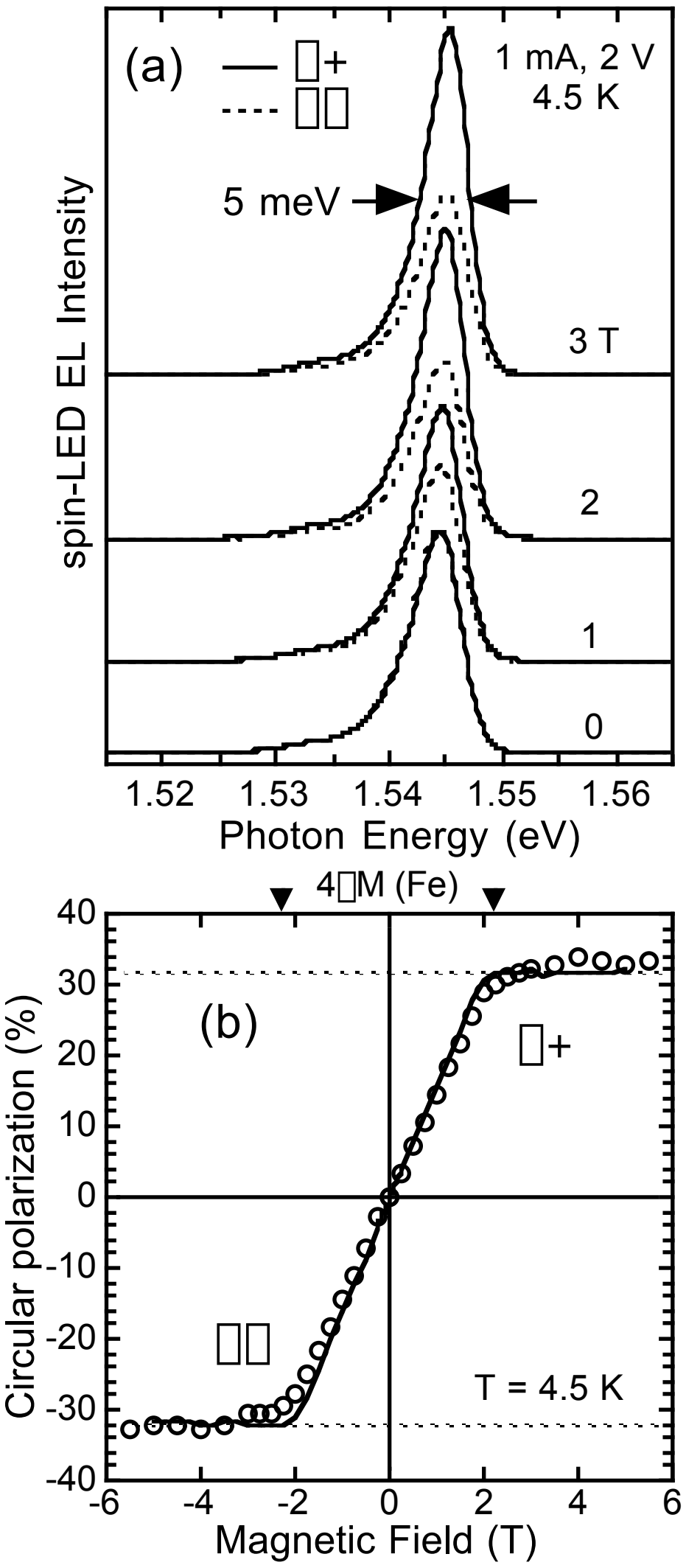

FIG. 2. (a) EL spectra from a spin-LED at selected applied magnetic fields analyzed for σ+ (solid) and σ– (dashed) polarization. (b) QW polarization as a function of applied magnetic field. The solid line is the hard axis magnetization of the Fe film as measured with SQUID magnetometry.

The spin polarization of the electron population in the GaAs QW, $P_{QW}$, produced by injection from the Fe Schottky contact is determined from the circular polarization of the surface-emitted electroluminescence (EL), $P_{circ}$ and the associated quantum selection rules.[18] Figure 2(a) shows the EL spectra from a spin-LED for applied magnetic fields of zero to 3 T. The applied field and observed light are along the surface normal (Faraday geometry). The spectra exhibit a single predominant feature due to heavy hole exciton recombination[18] with a full width at half maximum of ~ 5 meV. For each field, the EL is analyzed for positive (σ+) and negative (σ–) helicity circular polarization. At zero field, these components are coincident ($P_{circ} = 0$) because the Fe easy magnetization axis (and carrier spin orientation) lies in-plane. By 3 T, the Fe magnetization is saturated out-of-plane, and the two components exhibit a significant difference in intensity. Figure 2(b) shows the behavior of $P_{circ} = P_{QW}$[18] (open circles) vs. the out-of-plane magnetic field. $P_{QW}$ tracks the hard axis magnetization of the Fe film (solid line), demonstrating that the polarization originates from the Fe contact, and saturates at a value of 32% when the Fe magnetization is fully out-of-plane. This value, appreciably higher than reported previously,[7,8] demonstrates significant electrical spin injection across the Schottky barrier.

To determine the transport mechanism responsible for such large spin injection across the Fe/AlGaAs interface, we have examined the *I-V* characteristics in detail with standard ac and dc techniques. Attempts to fit *I–V* curves with standard Schottky barrier equations yield ideality factors much greater than 1, indicating a process other than pure thermionic emission dominates. This is expected because the structure was not designed as a conventional Schottky barrier. Therefore, we apply the "Rowell criteria."[12,15] which were originally developed to test whether single-step tunneling was the dominant transport mechanism for MIM structures.

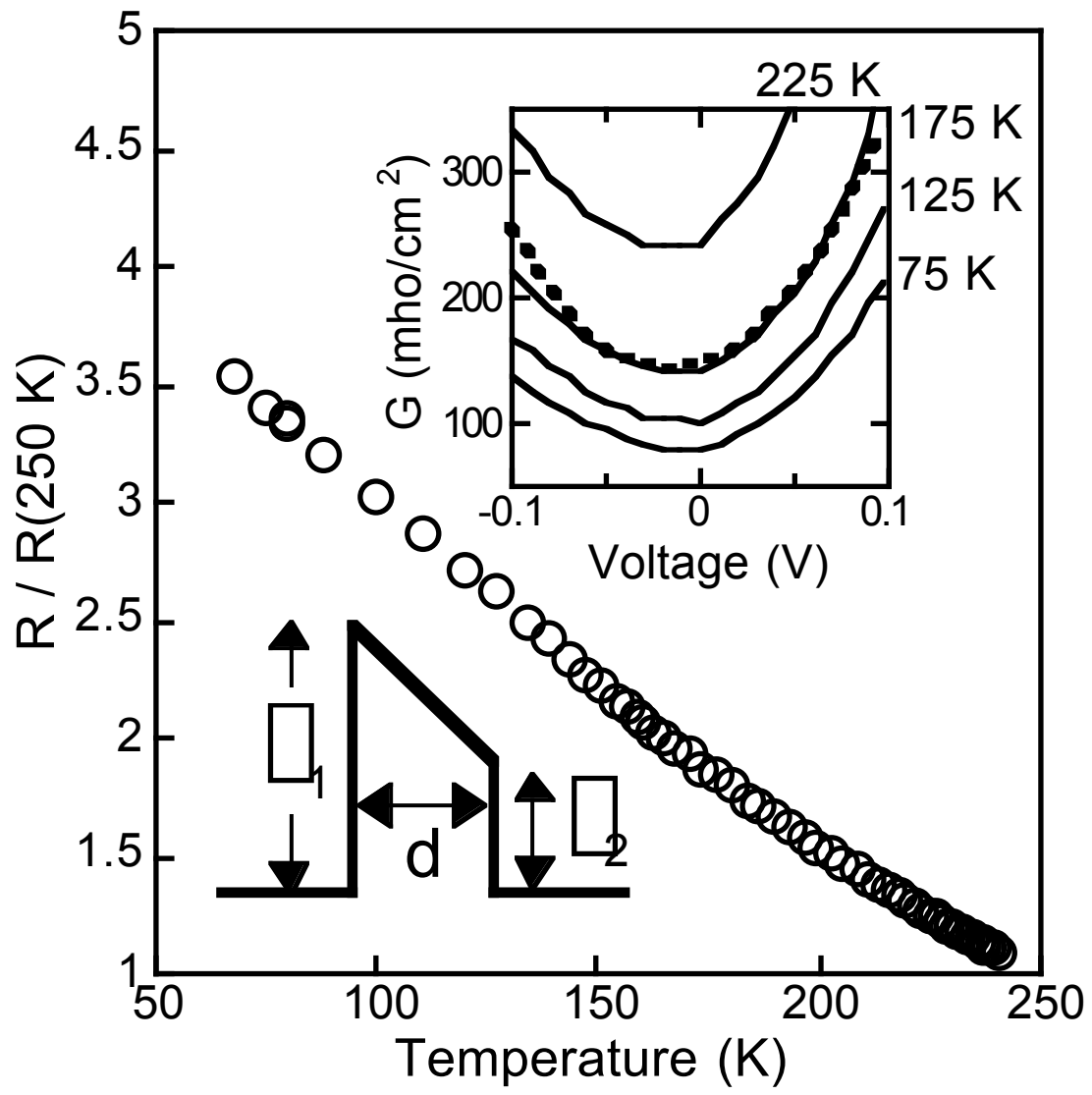

FIG. 3. Normalized ZBR as a function of temperature for an Fe/AlGaAs Schottky barrier contact. Inset is a series of conductance curves taken at different temperatures. The dotted line in the inset is a representative fit to the data. Parameters for the fitting are defined in the schematic.

There are three Rowell criteria. The first – the conductance ($G = dI/dV$) should have an exponential dependence on the thickness of the barrier – cannot be readily applied in this case due to the non-rectangular shape of the barrier and variations of the barrier width with bias. The second criterion states that the conductance should have a parabolic dependence on the voltage and can be fit with known models, e.g. a Simmons (symmetric barrier)[19] or Brinkman, Dynes and Rowell (BDR) model (asymmetric barrier).[20] The inset in figure 3 shows *G–V* data at a variety of temperatures, and a representative fit (dashed line) using the BDR model.

Parameters of this model are defined in the diagram. Fits to the data at seven different temperatures yield an average barrier thickness of d=29 Å, and barrier heights of $\phi_1$=0.46 eV and $\phi_2$=0.06 eV. The large potential difference between the two sides of the barrier is physically consistent with a triangular Schottky barrier tunnel junction. Although $\phi_1$ is lower than might be expected for an Fe/GaAs interface, image force lowering of the barrier due to the highly degenerate nature of the AlGaAs can lead to reduction of the barrier by more than 0.3 eV.[17] The "goodness" of the fits, energy range considered, and deviation of the fit parameters from "known" physical characteristics of the barrier are typical of similar treatments in the literature.[11,12] Therefore, we conclude that the second Rowell criterion is satisfied.

While the first two criteria are routinely invoked as proof of tunneling, it has been argued that *neither* can reliably distinguish tunneling from contributions due to spurious effects such as pinholes.[12,13,14] It has been shown that the *G–V* data can be fit with reasonable parameters even when tunneling was not the dominant transport path.[12] Jönsson-Åkerman, *et al.* have presented convincing evidence that the *third* Rowell criterion is a definitive confirmation of tunneling.[12] This criterion states that the zero-bias resistance (ZBR) should exhibit a weak, insulating-like temperature dependence. ZBR data are shown in figure 3 as a function of temperature, and clearly exhibit such a temperature dependence. Thus we conclude that the third Rowell criterion is also satisfied, confirming that single-step tunneling is the dominant conduction mechanism.

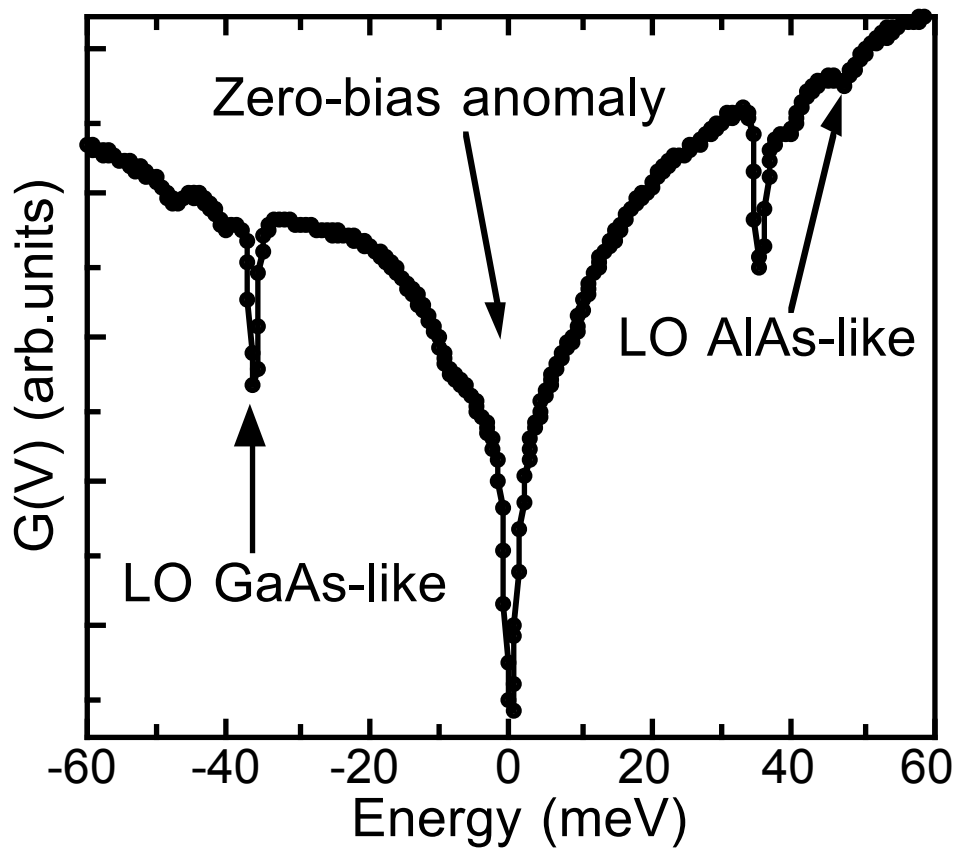

FIG. 4. (a) Conductance vs. applied voltage at 2.7 K. A zero-bias anomaly as well as phonon peaks attributed to GaAs-like and AlAs-like LO phonons are clearly visible.

To corroborate this conclusion, conductance measurements were performed at low temperatures in an effort to observe phonon modes of the AlGaAs barrier. [21] For transport across a diffusive contact, electron energies can never reach a value higher than a few kT above $E_f$, regardless of the applied bias, and phonon modes are not observed. A tunnel barrier enables transport of electrons at higher energies and a corresponding spectroscopy[21] – the observation of such features then provides further proof for tunneling.

The conductance spectrum for an $Fe/Al_{0.1}Ga_{0.9}As$ sample at 2.7 K is shown in figure 4, and exhibits two distinct features between 30 and 50 meV. The energy axis was scaled based on an independent measure of the series resistance which was of the same order as the device resistance. In AlGaAs, two sets of longitudinal-optical (LO) phonon modes are present, one GaAs–like and the other AlAs–like, with energies of 36 and 45 meV, respectively. The observed features agree very well with these nominal values, and are labeled accordingly. In addition, the relative intensities of the GaAs and AlAs phonon interactions are positively correlated with the relative Ga:Al content.[22] The observed features exhibit an intensity ratio of ~ 10:1, further confirming their identity.

At the lowest temperatures, the *G–V* data have a pronounced feature at zero bias. Zero bias anomalies are generally observed in semiconductor tunneling devices.[21] Although poorly understood, they have been attributed to inelastic scattering effects arising from acoustic phonons and barrier defects, and are closely associated with tunneling. The observation of such a feature in these data provides further evidence for tunneling.

In summary, we have observed a net electron spin polarization of 32% in a GaAs QW due to electrical spin injection from an Fe/AlGaAs reverse-biased Schottky contact. Application of the Rowell criteria demonstrates that single step tunneling is a significant transport mechanism. The conductance data show clear phonon signatures corresponding to GaAs-like and AlAs-like LO phonon modes, and a pronounced zero-bias. These observations provide conclusive evidence that tunneling is the dominant transport mechanism enabling significant spin injection across the metal / semiconductor interface.

This work was supported by the DARPA SpinS program, ONR, and NSF.